\documentclass{aa}  
\usepackage{graphicx}
\usepackage{txfonts}
\usepackage{natbib}
\bibpunct{(}{)}{;}{a}{}{,}   
\def\mxb{MXB 0656-072}

\begin{document}
   \title{New insights into the Be/X-ray binary system MXB 0656-072}


    \author{E. Nespoli
          \inst{1,2}
        \and
          P. Reig\inst{3,4}
        \and
        A. Zezas\inst{4}  
          }

   \offprints{Elisa Nespoli}

   \institute{Observatorio Astron\'omico de la Universidad de Valencia, Calle Catedr\'atico Jose Beltran, 2, 46980 Paterna, Valencia, Spain
   \and
   European Space Astronomy Centre (ESA/ESAC), Science Operations Department, Villanueva de la Ca\~nada (Madrid), Spain\\
       \email{elisa.nespoli@uv.es}
                \and
             Foundation for Research and Technology -- Hellas, IESL, Voutes, 71110 Heraklion, Crete, Greece
             \and
             Physics Department, University of Crete, 710 03 Heraklion, Crete, Greece\\
             }


 
  \abstract
   {The X-ray transient \mxb\ is a poorly studied member of high-mass X-ray binaries. Based on the transient nature of the X-ray emission, the detection of pulsations, and the early-type companion, it has been classified as a Be X-ray binary (Be/XRB). However, the flaring activity covering a large fraction of a giant outburst is somehow peculiar. }  
   {Our goal is to investigate the multiwavelength variability of the high-mass X-ray binary \mxb.}
   {We carried out optical spectroscopy and analysed all \emph{RXTE} archive data, performing a detailed X-ray-colour, spectral, and timing analysis of both normal (type-I) and giant (type-II) outbursts from \mxb. }
   { {This is the first detailed analysis of the optical counterpart in the classification region (4000-5000 A). From the strength and ratio of the elements and ions, we derive an O9.5Ve spectral type, in agreement with previous classification. This confirms its Be nature.} 
     {The characterisation of the Be/XRB system relies on} Balmer lines in emission in the optical spectra, long-term X-ray variability, and the orbital period vs. spin period  {and EW(H$\alpha$)} relation. The peculiar feature that distinguishes the type-II outburst is flaring activity, which occurs during the whole outburst peak, before a smoother decay. We interpret it in terms of magneto-hydrodynamic instability.  {Colour and spectral analysis reveal a hardening of the spectrum as the flux increases.} We explored  the aperiodic X-ray variability of the system for the first time, finding a correlation of the central frequency and $rms$ of the main timing component with luminosity, which extends up to a ``saturation'' flux of $1\times10^{-8}$ erg cm$^{-2}$ s$^{-1}$. A correlation between timing and spectral parameters was also found, pointing to an interconnection between the two physical regions responsible for both phenomenologies.}
   {}


\titlerunning{New insights into the Be/XRB system MXB 0656-072}
   \maketitle
%

\section{Introduction}

Be/X-ray binaries (Be/XRBs) constitute a sub-class of high-mass X-ray
binaries (HMXBs) in which the companion is a Be star, \emph{i.e.} a
non-supergiant fast-rotating OB-star that during its life has shown at some
point spectral lines in emission \citep[see][for a recent review]{rei11}. They are also characterised by infrared
excess, which means that they are brighter in the IR {than their non-emitting counterparts of the same spectral
type}. Both phenomena, emission lines and IR
excess, are thought to arise from a common cause, namely the presence of an
extended circumstellar envelope around the stellar equator, made up of
ionised gas that is expelled from the star in a way that is not yet
completely understood. This complex scenario is referred to as the \emph{Be
phenomenon} \citep{por03,eks08}. 

When a Be star is part of an X-ray binary, the system is
usually transient, and the compact object is virtually always a pulsar,
with typical spin periods ranging between 1--10$^{3}$ s. Be/XRBs are
characterised by high variability on a wide range of both time scales (from
seconds to years) and wavelengths, although the fastest variability is
observed in the X-ray band. For longer periods, the variability is apparent 
in both high-energy and low-energy wavelengths, and is attributed to major
changes in the circumstellar disc structure. The complexity of the dynamics
of the Be phenomenon and its relation with the accretion onto the compact
object clearly require a multiwavelength approach in the study of these
systems.

Even if its phenomenology is entangled and multi-faceted, the long-term X-ray variability in Be/XRBs is traditionally described by a classification
into two types of outbursts. Type-I (or normal) outbursts are periodic or
quasi-periodic events, occurring in correspondence (or close) to the
periastron passage of the neutron star. They are generally short, with a
typical duration of 0.2--0.3 $P_\mathrm{orb}$, and show luminosities
$L_\mathrm{X}=10^{36}$--$10^{37}$ erg s$^{-1}$. Type-II (or giant)
outbursts are unpredictable, long (one or more orbital periods), and
bright events, with typical X-ray luminosities of
$L_\mathrm{X}=10^{37}$--$10^{38}$ erg s$^{-1}$, corresponding to up to the
Eddington luminosity for a neutron star. The presence of quasi-periodic
oscillations (QPOs) in some systems would support the suggestion of the
formation of an accretion disc around the neutron star during type-II outbursts \citep[see for
instance][]{mot91,hay04}. 

The transient X-ray binary MXB 0656-072 was discovered by \emph{SAS-3} in
September 1975, when a flux density of 80 mCrab was reported \citep{cla75},
and subsequently observed twice in 1976 by \emph{Ariel V} at 50 (March 19)
and 70  (March 27) mCrab, respectively \citep{kal76}. These intensities
would convert into an X-ray luminosity of $\sim$2--$3\times10^{36}$ erg s$^{-1}$, assuming
a distance of 3.9 kpc \citep{mcb06}. Therefore, they would correspond to the
typical X-ray luminosity range for type-I outbursts.

Although the discovery of the source dates back to more than 35 years ago,
very little is known about the system. \mxb\ was only catalogued as an HMXB 
in 2003 after extended re-brightening, when its optical counterpart was
identified and classified as an O9.7Ve star \citep{pak03}, and a pulsed
period of 160.7s detected \citep{mor03}. The energy spectrum of the source
showed a cyclotron resonant energy feature (CRSF) at a central energy of
$\sim$33 keV \citep{hei03}. This event was classified as a type-II
outburst.

\citet{mcb06} monitored the source over the 2003 major outburst and studied the
average X-ray spectrum during the peak of the outburst and the change in
the spin period. They did a pulse phase-resolved analysis and found
that the width of the CRSF varied with pulse phase, being wider during the
pulse decline. However, the energy of the CRSF did not change with pulse
phase. A cyclotron line at $\sim$ 33 keV implies a magnetic field strength
of $3.7 \times 10^{12}$ G.  {They also present a timing analysis of the pulsations, including
pulse profile dependence on energy and the changes in the spin period
throughout the outburst.}
 {Recently, \citet{yan12} have presented a correlated optical/X-ray analysis of the system during type-I outbursts, after discovering a 101.2 days orbital period. }

In this paper we analyse all {\it RXTE} data available for \mxb, which
include the major 2003 outburst and a series of type-I outbursts observed
in 2007-2008. We performed X-ray colour, spectral, and timing analysis of the giant outburst, and colour and spectral analysis of the normal outbursts.  {We  focus on the aperiodic variability and the
study of the broad-band noise. Moreover, we present here additional optical spectroscopy that allows us to refine and robustly justify the spectral classification on one side, besides providing a long-term correlated H$\alpha$/X-ray follow up of the system.}

\section{\emph{RXTE} observations and data reduction}

{\it RXTE} followed the source during the first giant outburst observed
since its discovery, starting in October 2003 for approximately three and a half months.
(MJD 52931--53033). Renewed activity of the source was detected by the \emph{RXTE}
in November 2007, lasting for one year (MJD 54419--54776), with
luminosities lower than in 2003. The 2007 X-ray variability consisted of a
series of four (quasi-)periodic flares, which are reminiscent of type-I outbursts.
Table 1 shows the observation log. The total net exposure amounted to 129.7
ks for the first event analysed here, and to 368.7 ks for the second one.

 \begin{table}
      \caption{\small{Journal of \emph{RXTE} observations.}}
      \begin{center}
        \centering
            \begin{tabular}{cccc}
             \hline
             \hline
             \noalign{\smallskip}
                N. of           & Proposal  & MJD  & On-source     \\
                pointings &  ID &            range &  time (ks) \\
             \noalign{\smallskip}
              \hline
             \noalign{\smallskip}
                28   &   80067 &   52931.8--52975.4   & 91.1 \\           
                 33      & 80430  &  52966.9-- 53033.3     &  38.6\\ 
                44      & 93032  &  54419.5-- 54748.6     &  179.6\\ 
                 123      & 93423  &  54449.1-- 54776.6     &  189.1\\ 
         \noalign{\smallskip}
            \hline
\end{tabular}  
         \end{center}
   \label{tab:obslog}
  \end{table}

We employed data from all the three instruments onboard \emph{RXTE}
\citep{bra93}, the All-Sky Monitor (ASM), the Proportional Counter Array
(PCA), and the High Energy X-ray Timing Experiment (HEXTE). The ASM incorporates
three wide-angle shadow cameras equipped with proportional counters with a
total collecting area of 90 cm$^{2}$. It works in the 2--10 keV energy range,
mapping the 80\% of the sky every 90 minutes. The PCA consists of five
proportional counter units (PCUs) with a total collecting area of 
$\sim$6250 cm$^{2}$ and operates in the 2--60 keV range, with a nominal energy
resolution of 18\% at 6 keV. The HEXTE comprises two clusters of four
NaI/CsI scintillation counters, with a total collecting area of 2 $\times$
800 cm$^{2}$, sensitive in the 15--250 keV band with a nominal energy
resolution of 15\% at 60 keV. Both the PCA and the HEXTE have a maximum
time resolution of $\sim$1$\mu s$

Data reduction was performed using HEASOFT version 6.9. An energy spectrum
was obtained for each pointing, after filtering out unsuitable data
according to the recommended criteria\footnote{Among which, elevation from
the Earth greater than 10$^\circ$ and pointing offset lower than
0.02$^\circ$; see PCA digest at
http://heasarc.gsfc.nasa.gov/docs/xte/pca$\_$news.html}, employing Standard
2 mode data from the PCA (PCU2 only) and Standard (archive) mode from the
HEXTE Cluster A, with a time resolution of 16s. The PCA and HEXTE spectra were extracted, background-subtracted, and dead-time corrected. For the PCA, the 3--30 keV energy range
was retained, while the HEXTE provided a partially overlapping extension
from 25 to 100 keV. In the case of the 2003 outburst, for each observation, the two resulting spectra were
simultaneously fitted with XSPEC v.~12.6.0 \citep{arn96}. For the 2007-2008 event, three average PCA spectra were extracted in  three luminosity ranges ($<1\times10^{-9}$, $1-2.5\times10^{-9}$, and $>2.5\times10^{-9}$ erg cm$^{2}$ s$^{-1}$ respectively) and fitted between 3--60 keV in order to constrain the CRSF. Then a spectrum for each pointing was fitted between 3--30 keV, fixing the CRSF parameters to the corresponding average ones. HEXTE 2007--2008 spectra were too faint to be employed. During the fitting, a systematic error of 0.6\% was added to the PCA spectra.

Power spectral density (PSD) was computed using PCA Event or Single$\_$Bit
data. We first extracted, for each observation, a light curve in the energy
range $\sim$3.5--17 keV (channels 8--39) with a time resolution of $2^{-6}$
s. The light curve was then divided into 128-s segments, and a fast Fourier
transform was computed for each segment. The final PSD was computed as the
average of all the power spectra obtained for each segment. These averaged
power spectra were logarithmically rebinned in frequency and corrected for
dead time effects according to the prescriptions given in \citet{now99}.
Power spectra were normalised such that the integral over the PSD is equal
to the squared fractional $rms$ amplitude, according to the so-called
$rms$-normalisation \citep{bel90,miy91}.

\section{Optical observations}

\begin{table*}
\begin{center}
\caption{Log of the optical observations.}
\label{optobs}
\begin{tabular}{lcccccc}
\hline \hline \noalign{\smallskip}
Date &JD &Telescope &Grating&Wavelength &EW(H$\alpha$) &EW(H$\beta$)    \\
&(2,400,000+) & &(l/mm) &range (\AA) &(\AA) &(\AA)        \\
\hline \noalign{\smallskip}
14-11-2009$^*$ &55150.41 &FLWO &600 &4760-6760 &$-18.4\pm1.0$ &$-2.98\pm0.11$    \\
12-01-2010 &55209.37 &FLWO &600 &4730-6730 &$-20.9\pm1.5$ &$-3.43\pm0.08$    \\
16-01-2010 &55213.21 &FLWO &600 &4740-6740 &$-21.2\pm0.8$ &$-3.60\pm0.09$    \\
30-09-2010 &55470.60 &SKO &1301 &5300-7300 &$-10.8\pm0.4$ &--        \\
30-10-2010 &55501.01 &FLWO &1200 &6200-7200 &$-11.9\pm0.6$ &--        \\
29-11-2010 &55530.83 &FLWO &1200 &6200-7200 &$-11.6\pm0.6$ &--        \\
03-10-2011 &55838.98 &FLWO &1200 &6200-7200 &$-12.2\pm0.7$ &--        \\
01-11-2011 &55867.88 &FLWO &1200 &6200-7200 &$-13.0\pm0.7$ &--        \\
08-11-2011 &55873.60 &SKO &2400 &3940-5040 &-- &$-2.23\pm0.07$ \\
23-11-2011 &55889.93 &FLWO &1200 &6200-7200 &$-14.1\pm0.8$ &--        \\
31-11-2011$^*$ &55927.71 &FLWO &1200 &6200-7200 &$-15.4\pm0.8$ &--        \\
19-01-2012 &55946.79 &FLWO &1200 &6200-7200 &$-15.3\pm0.8$ &--        \\
22-01-2012$^*$ &55927.71 &FLWO &1200 &6200-7200 &$-15.2\pm0.8$ &--        \\
\hline \hline \noalign{\smallskip}
\multicolumn{7}{l}{$*$: Average of two measurements.}\\
\end{tabular}
\end{center}
\end{table*}

Optical spectroscopic observations of the companion star to MXB 0656-072
were performed using the Fred Lawrence Whipple Observatory at Mt. Hopkins
(Arizona, USA) and the 1.3-m telescope from the Skinakas observatory
(Crete, Greece). Table~\ref{optobs} gives the log of the observations. 

The 1.3m telescope of the Skinakas Observatory (SKO) was equipped with a
2000 $\times$ 800 ISA SITe CCD and a 1302 l mm$^{-1}$ grating (on 30
September 2010) and 2400 l mm$^{-1}$ (on 8 November 2011), giving a nominal
dispersion of $\sim$ 1 \AA/pixel and $\sim$ 0.5 \AA/pixel, respectively. We
also observed \mxb\ in queue mode with the 1.5-m telescope (FLWO) at
Mt. Hopkins (Arizona) and the FAST-II spectrograph plus FAST3 CCD, a
back-side illuminated 2688 $\times$ 512 UA STA520A chip with 15 $\mu$m
pixels.  Spectra of comparison lamps were taken before each exposure to account for small variations in the wavelength calibration during
the night. To ensure a homogeneous processing of the spectra, all of them
were normalised with respect to the local continuum, which was rectified to
unity by employing a spline fit. 

 {We measured calibrated photometry of the optical counterpart with dedicated observations for the first time}, performed from the 1.3-m telescope of
the Skinakas Observatory on 2 November 2010 (JD 2,455,503.6). \mxb\ was
observed through the Johnson B, V, R, and I filters. The telescope was
equipped with a 2048 $\times$ 2048 ANDOR CCD with a 13.5 $\mu$m pixel size.
Standard stars from the Landolt list \citep{lan09} were used for the
transformation equations. Reduction of the data was carried out in the
standard way using the IRAF tools for aperture photometry. The resulting
magnitudes are: $B=13.25\pm0.02$ mag, $V=12.25\pm0.02$ mag, $R=11.63\pm0.02$ mag, and 
$I=10.97\pm0.02$ mag. 

\section{Spectral class}

The only report of the spectral type of the massive companion in MXB
0656-072 is given by \citet{pak03}, who suggest an O9.7V spectral
class.  {Blue- and red-end optical spectra covering the period 2005-2009 are also presented in \citet{yan12}.}

\begin{figure}
\resizebox{\hsize}{!}{\includegraphics{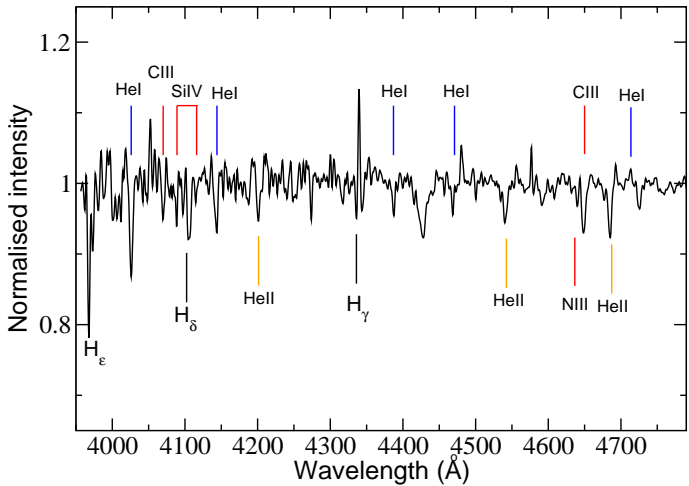} } 
\caption[]{The optical spectrum of the optical counterpart to MXB 0656-072
in the 4000-4800 \AA\ region. The identified lines correspond to He II
$\lambda$4200, $\lambda$4541, $\lambda$4686, He I $\lambda$4026, $\lambda$4144, $\lambda$4387, $\lambda$4471, $\lambda$4713, 
C III $\lambda$4070, $\lambda$4650, N III $\lambda$4640, Si IV
$\lambda$4089, $\lambda$4116, and the hydrogen lines of the Balmer series between H$_{\beta}$ and H$_{\epsilon}$. A Gaussian smoothing filter ($\sigma=1$) was
applied to reduce the noise.}
\label{speclass}
\end{figure}

Figure \ref{speclass} shows the optical spectrum of \mxb\ in the
region 4000-4800 \AA\ from the Skinakas observatory. This spectrum is the average of three spectra
obtained with a total exposure time of 7200s each. The distinct
presence of He II lines indicates an O-type star, while the presence of He I
lines implies that the spectral type must be later than O8. Some He I lines,
such as $\lambda$4713 and $\lambda$4921, appear (partially) in emission and
cannot be separated out from the continuum. The ratio \mbox{C III} $\lambda$ 4650
to He II $\lambda$4686 is close to 1, which agrees with an O9-O9.5 type
\citep{wal90}. The ratio He II $\lambda$4200/He I $\lambda$4144 allows us
to distinguish between these two close sub-classes \citep{wal71}. This ratio
is approximately 1 in O9 stars and lower than 1 in O9.5 stars. In MXB
0656-072 this ratio appears to be slightly lower than 1, favouring the
later type classification. Also,  since the strength of the He I $\lambda$4144 might be diminished by the emission from the circumstellar disc, the
O9.5 class appears to be more likely. On the other hand, given the
relatively low S/N in this part of the spectrum and the uncertainty
introduced in the definition of the continuum during the normalisation, an
O9 spectral type cannot be completely ruled out. 

As for the luminosity class, the strong He II $\lambda$4686 absorption
accompanied by weak N III $\lambda$4634-4640-4642 clearly indicates a
main-sequence star, and so does the strength of Si IV $\lambda$4089 in
comparison with that of He I $\lambda$4144. We conclude that the optical
counterpart to \mxb\ is an O9.5V star\footnote{The interpolated
class O9.7 suggested by \citet{pak03} is mainly used for supergiants. The
primary defining criterion for the O9.7 type is He II
$\lambda$4541$\approx$Si III $\lambda$4552 \citep{wal90}. According to
\citet{wal71}, it does not appear that this subdivision is useful at the
lower luminosities (class II and below) because the Si III lines are too
weak.}.

In addition to He I lines, the Balmer series of hydrogen lines are strongly
affected by emission. Even H$\epsilon$ appears to be filled-in with
emission. Columns 6 and 7 in Table~\ref{optobs} give the equivalent width
of the H$\alpha$ and H$\beta$ lines. A long-term decrease in the strength
of these two lines is observed.

 \begin{figure}
   \centering
 \includegraphics[width=8.5cm]{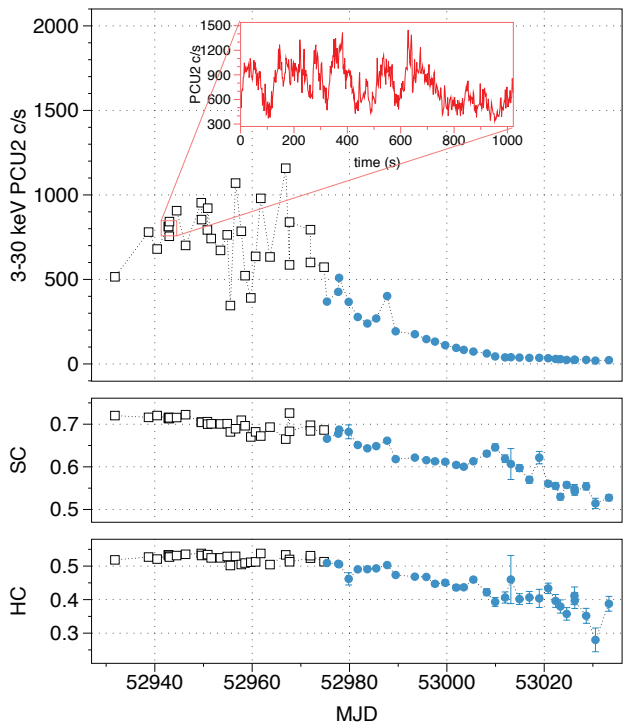}   
 \caption{PCA light curve and colour behaviour during the 2003 outburst. In the inset, 2s resolution light curve for one pointing during the flaring phase.}
     \label{fig:lc}
  \end{figure}

\section{Results}

We analysed all the observations of \mxb\ in the {\it RXTE} archive. We
separated the observations into two intervals corresponding to two
significant events. The first interval started on 20 October 2003 (MJD
52932) and covered a total of 101.5 days. This interval includes a giant
(type II) outburst. Assuming a distance of 3.9 kpc \citep{mcb06}, the
maximum 3--30 keV luminosity during the outburst was $L_\mathrm{X}=3.7
\times 10^{37}$ erg s$^{-1}$, registered at  MJD$\sim$52966.9. The second
interval covered the period  27 November 2007 to 11 November 2008 (MJD
54419.5--54776.6) and includes a series of minor (type I) outbursts. The
peak X-ray luminosity of these outbursts was $L_\mathrm{X}=1.37\times
10^{37}$ erg s$^{-1}$.

\subsection{Type II outburst}

\subsubsection{Colour analysis}

The PCA light curve and colour behaviour during the 2003 outburst is
presented in Fig.~\ref{fig:lc}. The outburst showed strong flaring
behaviour during the peak phase, followed by smoother decay. Each point
corresponds to an \emph{RXTE} pointing and is directly obtained from PCU2
count rate. Different symbols mark the different outburst phases, the
flare-like phase (open squares), and the smoother decay (filled circles).
Error bars are the same size as the points whenever they do not appear
in the plots. The X-ray colours were defined as follows, soft colour (SC):
7--10 keV / 4--7 keV; hard colour (HC): 15--30 keV / 10--15 keV.  The two
colours follow identical patterns during the outburst, both correlating
with flux. In the inset, a zoomed-in view is shown, presenting the light curve for one PCA pointing, with a 2s time resolution.
From the inset it is clear that the flare-like activity displayed during the outburst is large-scale behaviour that in fact corresponds to variability on various time scales, if investigated at higher time resolution: besides the $\sim$160s pulse period, slower and faster changes in intensity  are clearly detected in the 2s resolution light curve.

The amplitude of change in count rate in both colours is more than twice in the decay ($\sim$0.15) compared to the flare phase ($\sim$0.05, see Fig.~\ref{fig:lc}). 
Also, the colour values are higher larger during the flares, indicating a harder
spectrum.

\subsubsection{Spectral analysis}

We fitted energy spectra with a continuum constituted by a photo-absorbed
power law with a high-energy exponential cutoff, modified by a Gaussian line
at $\sim$6.5 keV with a fixed 0.5 keV width to account for Fe K$\alpha$
fluorescence. A CRSF at an average central energy of $~37$ keV was detected
in high-flux observations only, above $L_\mathrm{X}=1.6 \times 10^{37}$ erg
s$^{-1}$, \emph{i.e.} above $0.4 \times L_\mathrm{Xmax}$.  

In addition, \mxb\ shows significant 
residuals at $\sim$11 keV. They were fitted out by means of a Gaussian
absorption-like profile, which allowed acceptable fits, passing from
$\chi^{2}\sim$107 for  70 DOF to
$\chi^{2}\sim$85 for 67 DOF, for a typical high-flux observation. This
component is found at almost constant energy across the spectra, with a
weighted mean value of 11.68$\pm$0.05 keV, and was consistently reported
also by \citet{mcb06} {for the giant outburst and by \citet{yan12} for the normal ones}. We found that this feature is only necessary in spectra where the
CRSF is present as well. Its origin is uncertain. 

\begin{figure}
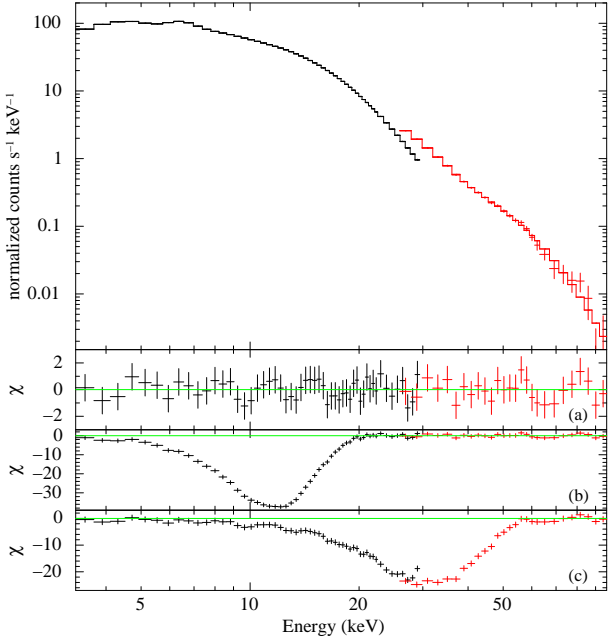

\begin{tabular}{ll}
\vspace{-4.7mm}
\includegraphics[bb= 80 14 522 776, clip,angle=-90,width=8cm]{specfit.ps} \\
\vspace{-8.7mm}
\includegraphics[bb=400 14 522 776,clip,angle=-90,width=8cm]{chi_a.ps} \\
\vspace{-8.7mm}
\includegraphics[bb=361 14 522 776,clip,angle=-90,width=8cm]{chi_b.ps} \\
\includegraphics[bb=361 14 581 776,clip,angle=-90,width=8cm]{chi_c.ps} \\
\end{tabular}
\caption[]{ {Typical spectrum for an observation during the flaring phase of the giant outburst (obsid: 80067-11-03-05, $L_\mathrm{x}$=2.8$\times10^{37}$ erg s$^{-1}$): in the upper panel, PCA and HEXTE data points and corresponding best fit are reported; below: residuals for the best fit (a), the fit excluding the absorption feature at $\sim$11 keV (b), and the one excluding the CRSF (c), respectively.}}
\label{x-ray-spec}
\end{figure}

{Figure \ref{x-ray-spec} shows a typical spectrum for a high-flux observation, with the corresponding residuals in the case of the best fit (a), the fit excluding the absorption at $\sim$11 keV (b), and the one excluding the CRSF (c) from the model.}

   \begin{figure}
  \centering
   \includegraphics[width=8.5cm]{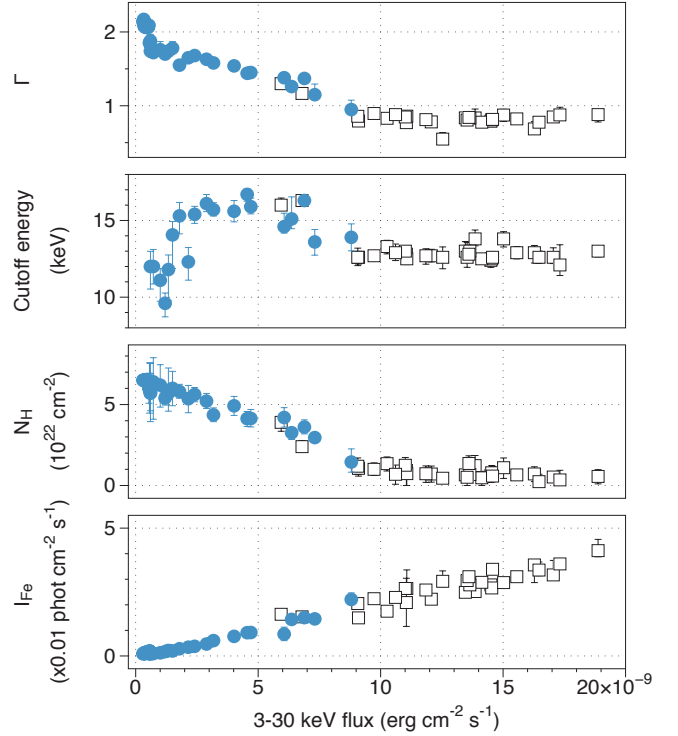}   
  \caption{Evolution of the main spectral parameters during the 2003 outburst.
   From the upper panel: photon index, cutoff energy, hydrogen column density,
    and Fe line intensity.}
      \label{fig:sp}
   \end{figure}

The best-fit main spectral parameters are shown in Fig.~\ref{fig:sp} as a
function of the calculated 3--30 keV flux. Different marks and colours
identify the two different phases of the outburst, the flaring and the
smooth one. Points separated in time and corresponding to
different phases behave in a similar way, which only depends on flux.  The
power-law photon index decreases with X-ray flux, confirming the result
from the colour analysis that as the X-ray flux increases the spectrum
becomes harder (see also Fig.~\ref{fig:lc}). The strength of the iron fluorescence line nicely
correlates with flux, revealing an increase in the reprocessed material as
the flux increases, as expected. The central energy of the Fe line remained
fairly constant, with a mean value of 6.49$\pm$0.06 keV, and a
mean equivalent width (EW) of  0.34$\pm$0.07 keV.

The CRSF was almost constant during the outburst, with the following
weighted mean values: $E_\mathrm{c}=36.8\pm0.4$ keV, $\sigma=9.1\pm0.4$.
These values are not compatible with the best-fit values by \citet{mcb06},
who found the CRSF at a central energy of $32.8^{+0.5}_{-0.4}$ keV. The
discrepancy may be because \citet{mcb06} obtained one spectral
fit from a spectrum retrieved by summing up all the spectra from the flare
phase of the outburst (from MJD 52932 to MJD 52964). In fact, when choosing
four individual observations of that interval, the cyclotron line in the
spectra of those four observations showed line energies above 33 keV
\citep[see Fig. 3 in][]{mcb06}. Our best-fit value for the central energy
agrees with the one firstly reported by \citet{hei03}, of
$E_\mathrm{c}=36\pm1$ keV.

All the observed correlations, except for the iron line strength, are significant up to a ``saturation'' flux of $\sim$10$^{-8}$ erg cm$^{-2}$ s$^{-1}$, which corresponds to L$_\mathrm{X}=1.8\times10^{37}$ erg s$^{-1}$. This luminosity roughly coincides with the large-amplitude flaring phase.

Because of the flaring activity, it is difficult to identify the peak of
the outburst. The average 3-30 keV X-ray luminosity in the time interval
MJD 52940--52970 is $L_\mathrm{X}=2.5 \times 10^{37}$ erg s$^{-1}$,
although a maximum of $L_\mathrm{X}=3.7 \times 10^{37}$ erg s$^{-1}$ was
obtained on MJD 52966.9. The lowest luminosity corresponds to the last
point of the decay with $L_\mathrm{X}=1\times 10^{36}$ erg s$^{-1}$.

\subsubsection{Timing analysis}

In this work, we focus on the aperiodic variability of the system. For a study of the X-ray pulsations, see \citet{mcb06}. The neutron star spin frequency has a
fundamental peak at $\sim$6 mHz, below our PSD frequency range
(0.008--32 Hz). Thus, peaks derived from the neutron star's pulsations do
not appreciably distort the continuum in the power spectra.

We fitted each PSD with the sum of Lorentzian functions with the objective
of providing a unified phenomenological description of the timing behaviour
of the system during the outburst. We denote each component as
$L_\mathrm{i}$, and its characteristic frequency $\nu_\mathrm{max}$ as
$\nu_\mathrm{i}$. According to the definition in \citet{bel02}, this is the
frequency where the component contributes most of its variance per
logarithmic frequency interval, $\nu_\mathrm{max} = \sqrt{\nu_{0}^{2}+
(FWHM/2)^{2}}$, where $\nu_{0}$ is the centroid frequency and $FWHM$ is the
full width at half maximum of the Lorentzian function. In this work, we
always refer to characteristic frequencies $\nu_\mathrm{max}$.

   \begin{figure}
  \centering
   \includegraphics[width=8.5cm]{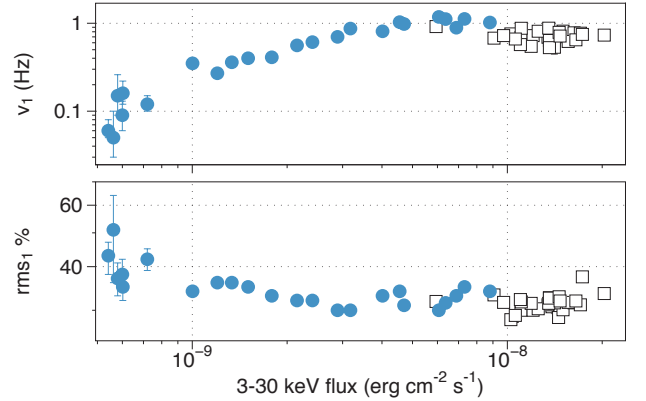}   
  \caption{Evolution of the timing parameters, characteristic frequency and 
  fractional $rms$, of the best constrained noise component, $L_{1}$, during the 
  2003 outburst.}
      \label{fig:tim}
   \end{figure}

   \begin{figure}
  \centering
   \includegraphics[width=8.5cm]{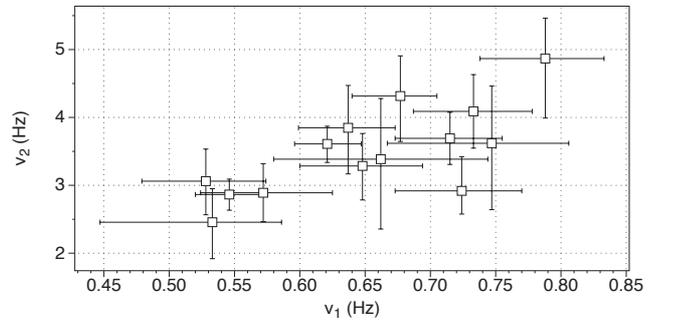}   
  \caption{Relation between the maximum frequencies of the $L_{1}$ and $L_{2}$ 
  components.}
      \label{fig:timcorr}
   \end{figure}

The low and middle-frequency noise
($L_0$ and $L_1$) is accounted for by zero-centred Lorentzians with a
characteristic frequency that is generally lower for low-flux pointings, and
higher for high-flux observations. The corresponding fractional $rms$ varies
during the outburst, in anti-correlation with flux. These components are the only ones required during all the
smooth decay phase (after \mbox{MJD 52975}) and part of the flaring phase of the outburst, up to $f_x\approx 1 \times 10^{-8}$ erg cm$^{-2}$ s$^{-1}$.  Beyond that luminosity, an additional component is necessary, $L_{2}$, whose characteristic frequency
 varies
in the range $\sim$1-5 Hz, whereas the fractional amplitude of variability
varies between 10\% and 20\%, without a clear dependence on flux.  Unlike $L_0$
and $L_1$, this component has, in general, a non-zero centroid frequency, and its average value for the $Q$-factor, defined as $Q=\nu_{0}/FWHM$, is $\sim$0.6, denoting a narrower feature compared to $L_{0}$ and $L_{1}$.
Figure~\ref{fig:tim} presents the evolution of the characteristic frequency
and $rms$ over the outburst for the $L_{1}$ component, the best constrained
one. Up to  $f_x\approx 1 \times 10^{-8}$ erg cm$^{-2}$ s$^{-1}$, as the flux increases the characteristic frequency increases, whilst after that luminosity it remains approximately constant. The $rms$ variability shows an opposite trend, decreasing as the flux increases, and then saturating. Figure~\ref{fig:timcorr} shows the relation between the $L_{1}$ and
$L_{2}$ characteristic frequencies; although within some scattering, the two frequencies follow a correlated trend in all their range 
of variation.

   \begin{figure}
  \centering
   \includegraphics[width=8cm]{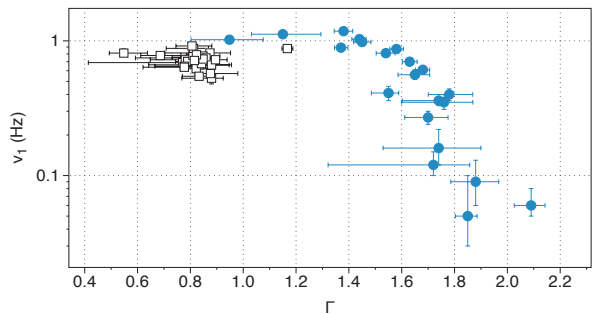}   
  \caption{Relation between the photon index and the characteristic frequency of the $L_{1}$ broad-noise component.}
      \label{fig:corr_sptim}
   \end{figure}

We studied the correlated spectral/timing behaviour, and found that during the very last part of the flaring phase, and the whole decay, the photon index $\Gamma$ and the central frequency of the main timing component, $\nu_{1}$, vary in an anti-correlated way (Fig.~\ref{fig:corr_sptim}). During the flares the $L_{1}$ frequency shows only slight variation due to the appearance of $L_{2}$, so that no spectral/timing relation could be expected.

\subsection{Type I outbursts}

\begin{figure}
\includegraphics[bb= 15 30 740 530, clip, width=9cm]{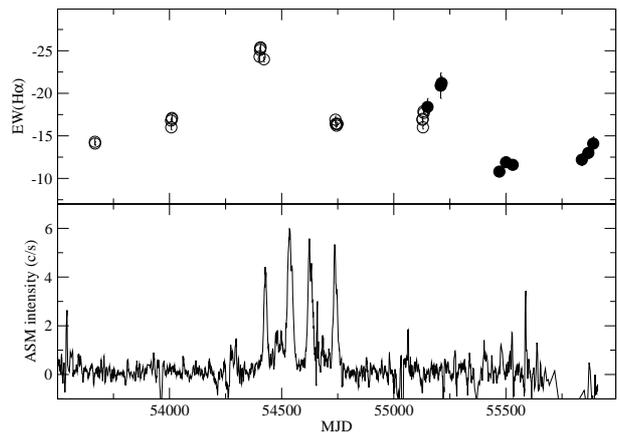} 
\caption[]{ASM light curve and H$\alpha$ equivalent width evolution. A
5-day running average was applied to the 1-day ASM light curve
to reduce the noise. Open circles correspond to observations by
\citet{yan12} and filled circles to our observations (see Table~\ref{optobs}).}
\label{typeI-ew}
\end{figure}

In addition to the major (type II) outburst reported in previous sections,
MXB 0656-072 underwent a series of four fainter outbursts between November
2007 and November 2008. The PCA began to monitor these outbursts at the end of the first one. All
outbursts exhibited similar peak luminosities ($L_{\rm peak}=1.2\times
10^{37}$ erg s$^{-1}$). This luminosity is about three times lower than that
of the November 2003 type-II outburst. 

We show in Fig.~\ref{typeI-ew} the simultaneous H$\alpha$ measurements and ASM light curve. The outbursts are separated by $\sim$100 days and in between the major peaks, other minor peaks are observed. The onset of these outbursts coincided with an
optical maximum brightness of the donor: at around MJD 54500
(February 2008) the equivalent width of the H$\alpha$ line seems to have
reached a maximum value of $\sim$25 \AA\ \citep{yan12}. 

   \begin{figure}
   \begin{tabular}{c}
   \includegraphics[bb= 15 30 740 530, clip, width=8cm]{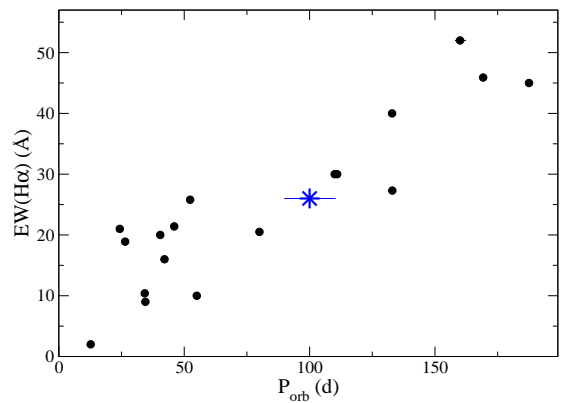} \\
   \end{tabular}
      \caption{ $P_{\rm orb}-EW(H\alpha$)
      diagram. The star symbol marks the position of \mxb. }
         \label{diagrams}
   \end{figure}

In our X-ray spectral analysis, in order to best constrain the CRSF, hardly detectable at very low flux, we extracted three average spectra in three ranges of luminosities and fitted them between 3--60 keV. The model employed was the same as for type-II outburst, and all the components were needed to obtain acceptable fits at in all the three spectra. Spectral parameters are shown in Table~\ref{tab:spec-typeI}. Similarly to our analysis of the giant outburst, we also generated one
energy spectrum for each observing interval and studied the evolution of
the spectral parameters.  We fixed the CRSF parameters in each single spectrum to the ones obtained from the corresponding flux average spectrum, and fitted each one between 3 keV and 30 keV.  
We found that the spectral parameters follow similar trends to those seen during the type II
outburst (Fig.~\ref{fig:sp}). The photon index decreases as the flux increases, as in the
type II outburst, although with a steeper dependence
with flux. The iron line energy does not significantly vary, while its intensity, expressed as normalisation, tightly correlates with luminosity. The iron line width was fixed at 0.5 keV. The hydrogen column density generally anti-correlates with flux, while the cutoff energy does not show any smooth relation with luminosity, although it displays lower values at higher flux and vice versa.   
 
\begin{table}
\caption{Type-I outburst spectral analysis for average spectra at different flux ranges.}
\label{tab:spec-typeI}
\begin{tabular}{lccc}
\hline \hline \noalign{\smallskip}  
Spectral parameter		&low flux$^{a}$	& med. flux$^{b}$	& high flux$^{c}$ \\
\hline \noalign{\smallskip}
$\Gamma$					&1.08$\pm$0.08	&0.83$\pm$0.04	&0.49$\pm$0.03		\\{\smallskip}  
cutoff en. (keV)					&16.9$_{-2.3}^{+2.5}$	&14.4$_{-0.5}^{+0.9}$	&10.5$\pm$0.02		\\{\smallskip}  
pow. norm. (ph/keV/cm$^{2}$/s)	&0.11$_{-0.02}^{+0.04}$	&0.12$\pm$0.01	&0.115$_{-0.007}^{+0.004}$	\\{\smallskip}  
nH (10$^{22}$cm$^{-2}$)			&4.1$\pm$0.5	&2.8$\pm$0.3		&1.9$\pm$0.3			\\{\smallskip}  
E$_\mathrm{Fe}$ (keV)			&6.5$\pm$0.1	&6.45$_{-0.04}^{+0.05}$	&6.5$\pm$0.4			\\{\smallskip}  		
EW$_\mathrm{Fe}$				&0.06$\pm$0.02	&0.14$_{-0.8}^{+0.2}$	&0.17$_{-0.11}^{+0.22}$		\\{\smallskip}  
E$_\mathrm{10keV-gaus}$		&10.3$_{-0.5}^{+0.4}$	&10.6$_{-0.1}^{+0.3}$	&10.6$_{-0.1}^{+0.2}$\\{\smallskip}  
$\sigma_\mathrm{10keV-gaus}$	&6.5$_{-0.6}^{+0.9}$	&4.2$_{-0.5}^{+0.3}$	&2.9$_{-0.23}^{+0.24}$\\{\smallskip}  
$\tau_\mathrm{10keV-gaus}$		&15.8$_{-3.9}^{+6.2}$	&2.3$_{-0.5}^{+0.4}$	&1.05$_{-0.12}^{+0.16}$	\\{\smallskip}  
E$_\mathrm{cyc}$				&31.9$_{-2.7}^{+1.3}$	&35.6$_{-1.3}^{+2.1}$	&35.0$_{-1.0}^{+1.7}$	\\{\smallskip}  
$\sigma_\mathrm{cyc}$			&11.2$_{-0.9}^{+3.2}$	&8.5$_{-1.2}^{+1.4}$	&8.8$_{-1.3}^{+2.2}$	\\{\smallskip}  
$\tau_\mathrm{cyc}$				&58.0$_{-8.9}^{+13.2}$	&14.0$_{-3.7}^{+8.1}$	&13.0$_{-3.8}^{+7.8}$		\\
\hline \hline \noalign{\smallskip}
\multicolumn{4}{l}{$a$: $<10^{-9}$ erg cm$^{2}$ s$^{-1}$  }\\
\multicolumn{4}{l}{$b$: (1--2.5) $\times10^{-9}$ erg cm$^{2}$ s$^{-1}$ }\\
\multicolumn{4}{l}{$c$: $>2.5\times10^{-9}$ erg cm$^{2}$ s$^{-1}$}\\
\end{tabular}
\end{table}

\section{Discussion}

We have performed a detailed X-ray and optical analysis of the poorly
studied hard X-ray transient \mxb. All the available observational data
indicate that \mxb\ is a member of the class of massive X-ray binaries
known as Be/X-ray binaries. X-rays are produced in the vicinity of
the compact object, while the optical variability comes from the young and
massive companion. The detection of X-ray pulsations, the transient nature
of the X-ray emission, and the characteristics of the  X-ray spectrum
(power-law continuum modified by an exponential cutoff and the presence of
fluorescence iron line and cyclotron feature) are typical of neutron star
binaries. The observation of Balmer lines in emission favours the Be/XRB classification. Furthermore, the long-term
X-ray variability, consisting of giant (type II) and minor recurrent
outbursts (type I) are typical of Be/XRB. 

\subsection{Optical observations}

The optical counterpart to \mxb\ was classified as a O9.5Ve
star, {refining previous classification by \citet{pak03}}. Its spectrum is strongly affected by emission, with the first
three lines of the Balmer series (H$\alpha$, H$\beta$, and H$\gamma$) showing
an emission profile, while the next two (H$\delta$ and H$\epsilon$) are
partially filled in with emission. This extra emission is thought to arise from the
equatorial disc around the Be star. The picture described by optical spectroscopy is fully consistent with the so-called 
``Be-phenomenon'' and confirms that the system is a Be/XRB. 

 {From the analysis of type-I outbursts, \citet{yan12} find an orbital period of 101.2d.  Once the period is known, we can use two important relationships involving the orbital period of the
system, the $P_{\rm spin}-P_{\rm orb}$ \citep{cor86} and $P_{\rm
orb}-EW(H\alpha)$ \citep{rei97,rei11} diagrams, to support the orbital period found. In the first diagram \citep[see Fig. 6 in][]{yan12}, the source is clearly located in the region occupied by Be/XRBs; in the second one (Fig.~\ref{diagrams}), a $\sim$100d orbital period fits nicely in the expected  EW(H$\alpha$) vs. orbital period relation. 
 The $P_{\rm orb}-EW(H\alpha)$ correlation is a consequence of tidal
truncation of the Be star's circumstellar disc by the neutron star. 
Assuming that the equivalent width of the H$\alpha$ line, $EW(H\alpha)$,
provides a good measure of the size of the circumstellar disc \citep{qui97,tyc05}, the $P_{\rm orb}-EW(H\alpha)$
correlation indicates that systems with long orbital periods have larger
discs, while narrow orbit systems contain smaller discs. In short orbital
period systems, the neutron star prevents the formation of extended discs. As can be seen in Fig.~\ref{diagrams}, an orbital period of $\sim100$ days
agrees very well with the
maximum $EW(H\alpha)$ of $\sim -25$ \AA.}

\subsection{Type-II outburst}

The profile of the type-II outburst in \mxb\ is unusual among Be/XRB. Although
flaring behaviour has been seen in other systems \citep[\emph{e.g.,} EXO 2030+375, A0535+26,][]{klo11,cab08}, the profile
of the outbursts tends to be more symmetric and to have a smoother peak. In
\mxb, the flaring activity covers a substantial part of the outburst
(Fig.~\ref{fig:lc}). It was found that magneto-hydrodynamic instabilities at the inner edge of the accretion disc may produce oscillations in the accreting flow, possibly leading to the observed behaviour \citep{app91,pos08,dan10}. Especially, \citet{dan10} show that, if the accretion disc is truncated by the neutron star's magnetic field outside but close to the corotation radius (at which the Keplerian frequency in the disc equals the star's rotational frequency), then the accretion becomes time-dependent and takes the form of repeated bursts.  

We found correlated behaviour of both the SC and the HC with flux, which
corresponds to a general hardening of the spectra as the flux increases. The dependence of the SC on luminosity does, in appearance, contrast with the observed negative correlation in other sources.  \citet{rei08} analysed the spectral and timing properties of four Be/XRBs during giant outburst, identifying two \emph{source states}, a low-luminosity horizontal branch (HB) and a high-luminosity diagonal branch (DB). During the outburst, the sources spend most of the time in the DB, during which the SC anti-correlates with flux. The transition to the HB inverts this trend, and the SC starts correlates with it. The correlation between the SC and luminosity observed in \mxb\ thus reveals that the source never undergoes the transition to the DB, but remains in the HB during all the duration of the outburst.

The hardening of the spectrum at high flux is confirmed by spectral
analysis: from simultaneous spectral fitting of PCA and HEXTE spectra, we
found a clear anti-correlated behaviour of the spectral index with flux
(Fig.~\ref{fig:sp}) up to flux $\sim1.2\times10^{-8}$ erg cm$^{2}$
s$^{-1}$, or $L_\mathrm{X}=2.3\times10^{37}$ erg s$^{-1}$, when the trend
flattens. The spectral hardening with increasing flux can be interpreted with simple Comptonisation models: the X-ray emission from HMXBs is thought to arise in an accretion column, close to the neutron star's surface, as a result of Comptonisation of soft photons injected in the accretion flow from the NS thermal mound, by high-energy electrons of the accreting matter \citep{bec07}.   Increasing X-ray luminosity would mainly correspond to increasing mass accretion rate, which can be translated into averagely  increasing number of up-scattering collisions of photons with electrons, resulting in a harder spectrum. 

After 1A 1118-615 \citep{nes11}, \mxb\ is the second X-ray pulsar to show correlated spectral/aperiodic behaviour (Fig.\ref{fig:corr_sptim}). This translates into a necessary connection between the two physical regions responsible for producing the two phenomenologies, the accretion column on one side, where energy spectra arise, and the accretion disc on the other side where, according to the ``perturbation propagation'' model, the aperiodic variability is located \citep[][and references therein]{rev09}. In this model, in fact, the X-ray variability is caused by perturbations in the inner disc flow, at different radii. The two regions are physically separated, being the first close to the NS polar cap regions, and the second confined outside the magnetosphere, but our results on two systems show that they are somehow coupled. 

In general, the X-ray variability of \mxb\ during the
type-II outburst resembles that of 1A 1118-615 at many levels.  The colours behave exactly the same in the two sources (both correlate with flux). The spectral parameters have the same trend in the two sources, and in both cases the relation with flux ceases at some saturation luminosity. In \mxb\ this saturation is reached at 0.5$\times L_\mathrm{Xmax}$, while in 1A~1118-615 at 0.7$\times L_\mathrm{Xmax}$. Finally, the correlations between the two broadband timing components and between spectral and timing parameters are also observed in both sources, making them very similar during a type-II outburst, although no flaring activity is observed in  1A 1118-615.

A peculiar feature in the X-ray energy spectra of \mxb\ is an absorption-line-like profile at an average 
constant energy between 11--12 keV.  This component was investigated by \citet{cob01}, who detected it in the range 8-12 keV in the spectra of many X-ray pulsars. The feature was consistently observed at the same energy, irrespective the CRSF energy, and moreover, it was evident in some systems that do not display a cyclotron line. This made \citet{cob01} conclude that the component should not be a magnetic effect. The feature seems to be intrinsic to X-ray pulsars spectra, since it was observed with different instruments \citep[besides \emph{RXTE}, \emph{Ginga} and  \emph{Beppo}SAX,][]{mih95, san98}. In the case of \mxb, this component is only found in spectra where the CRSF is present as well, although no other relation could be established between the two features.

\section{Conclusions}

We presented a detailed X-ray and optical study of \mxb\ covering both types of X-ray variability observed in a Be/XRB, namely type-I and type-II outbursts. The major outburst is characterised by flare-like behaviour during its peak, followed by smoother decay. We interpreted the flaring activity as possibly due to magneto-hydrodynamic instabilities at the inner edge of the accretion disc. The colour and spectral analyses reveal a hardening of the spectra as the luminosity increases, which can be understood in the framework of the models for spectra production in X-ray pulsars. The analysis of aperiodic variability shows correlated behaviour of the timing parameters with flux, which translates into a correlation between spectral/timing features and can be interpreted as an interconnection between the two physical regions responsible for the two phenomenologies. All the X-ray behaviour during the type-II outburst resembles that of 1A 1118-615, although no such flaring activity was observed in that system. 
 {The spin period vs. EW(H$\alpha$) relation confirmed the orbital period proposed for the source. }
The full multiwavelength analysis corroborates the Be/XRB nature of the system. 

Further observations during major outbursts are needed in order to explore the nature of the flaring activity and the timing/spectral correlation detected in this work. Deeper comprehension of the interaction between the magnetosphere and the accretion disc is also necessary to explain the correlated behaviour of the spectral and aperiodic features.

\acknowledgements{EN acknowledges a ``VALi+d'' postdoctoral grant from the ``Generalitat Valenciana'' and was supported by the Spanish Ministry of Economy and Competitiveness under contract AYA 2010-18352. PR acknowledges support by the Programa Nacional de Movilidad de Recursos Humanos de Investigaci\'on 2011 del Plan Nacional de I-D+i 2008-2011 of the
Spanish Ministry of Education, Culture and Sport. PR also acknowledges
partial support by the COST Action ECOST-STSM-MP0905-020112-013371.
ASM quick-look results provided by the ASM/\emph{RXTE} team. }

\label{lastpage}

\bibliographystyle{aa}
\bibliography{mxb0656}

\end{document}